# Surface anchoring as a control parameter for stabilizing torons, skyrmions, twisted walls, fingers and their hybrids in chiral nematics

Jung-Shen B. Tai (戴榮身)[1] and Ivan I. Smalyukh[1,2,3*]

[1]Department of Physics, University of Colorado, Boulder, CO 80309, USA
[2]Materials Science and Engineering Program, Soft Materials Research Center and Department of Electrical, Computer and Energy Engineering, University of Colorado, Boulder, CO 80309, USA
[3]Renewable and Sustainable Energy Institute, National Renewable Energy Laboratory and University of Colorado, Boulder, CO 80309, USA

*Corresponding author: ivan.smalyukh@colorado.edu

*Abstract:* Chiral condensed matter systems, such as liquid crystals and magnets, exhibit a host of spatially localized topological structures that emerge from the medium's tendency to twist and its competition with confinement and field coupling effects. We show that the strength of perpendicular surface boundary conditions can be used to control the structure and topology of solitonic and other localized field configurations. By combining numerical modeling and three-dimensional imaging of the director field, we reveal structural stability diagrams and inter-transformation of twisted walls and fingers, torons and skyrmions and their crystalline organizations upon changing boundary conditions. Our findings provide a recipe for controllably realizing skyrmions, torons and hybrid solitonic structures possessing features of both of them, which will aid in fundamental explorations and technological uses of such topological solitons. Moreover, we discuss how other material parameters can be used to determine soliton stability, and how similar principles can be systematically applied to other liquid crystal solitons, and solitons in other material systems.

# I. INTRODUCTION

As originally introduced, Skyrme's topological solitons [1–4] describe sub-atomic particles with different baryon numbers in nuclear physics. However, similar topologically-protected solitonic structures also occur in many other branches of science, ranging from optics to cosmology [5–7]. In condensed matter, low-dimensional analogues of Skyrme's solitons (often called "baby skyrmions") recently attract a great deal of interest in chiral liquid crystals (LCs) and magnets, with the latter promising racetrack memory and various spintronics applications, which recently fueled an explosive growth of their studies [8–12]. There is also great deal of interest in studying solitons in ferromagnetic LCs recently, which are formed by colloidal dispersions of magnetically monodomain nanoparticles [13,14]. In ferromagnetic LCs, the polar ordering in magnetization allows for linear coupling and facile response to applied magnetic fields, enabling additional controls of soliton stability and topology [13–15]. External fields and thin-film confinement are widely used to control stability of such topologically nontrivial configurations in these condensed matter systems. In addition to the two-dimensional (2D) baby skyrmions, which are most widely known [8–12,16–19], studies of chiral LC systems revealed various torons, toron-skyrmion hybrids, hopfions, heliknotons, twistions and so on [15,16,20–26]. Similar configurations were also found in magnetic systems, though sometimes they were given different names. For example, localized structures of elementary torons initially observed in LCs were also found in chiral magnets, although there they were called both "torons" [27] and "monopole-antimonopole pairs" [28]. When later discovered in magnetic systems, the toron-skyrmion hybrids were referred to as "bobbers" [29], whereas twistions have been found only in LCs so far [22,23]. Surface boundary conditions were also a useful means of stabilizing structures like hopfions in both LCs and in noncentrosymmetric magnets, though they were demonstrated in both experiments and theoretical modeling in LCs [23,24] but only in numerical modeling in magnets so far [28,30,31]. Despite the well-documented importance of surface boundary conditions [16], their role in stabilizing different one-, two, and three-dimensional solitonic LC structures has not been systematically explored so far.

In this article, we demonstrate that surface anchoring boundary conditions in LCs allow for pre-defining stability of torons, baby skyrmions, or the topologically trivial unwound state. We numerically establish structural stability diagrams, which we validate by means of three-dimensional imaging of director fields and independent characterization of the strength of perpendicular surface boundary conditions. We show that, in addition to chiral interactions, confinement and surface interactions help overcome the constraints of Derrick-Hobart theorem [32] by enhancing stability of solitonic configurations. Since many techniques for controlling the surface anchoring have been developed by the LC display industry and since the toron and baby skyrmion structures can be readily realized under conditions matching those used in displays, we foresee that the topological solitons can gain not only fundamental physics importance, but also can find applications in new modes of displays and electro-optic devices. Moreover, we discuss how the surface anchoring-based means of controlling solitonic structures can be extended to other one- and three-dimensional topological solitons and related structures.

## II. NUMERICAL METHODS

### II.1. Free Energy Functional

To model the molecular alignment field $\boldsymbol{n}(\boldsymbol{r})$ of solitonic configurations in chiral LCs between glass substrates with homeotropic anchoring, both Frank-Oseen and Landau-de Gennes free energy functionals were used for energy minimization. Frank-Oseen free energy was used for vector fields and orientable director fields (director fields that can be dressed smoothly with vectors without introducing extra fictitious singularities) [33]. All solitonic structures discussed in this work can be modeled by minimizing the Frank-Oseen free energy, except for the cholesteric finger of the third type (CF-3), where Landau-de Gennes modeling is used to properly model singular half-integer defect lines within their structures.

In Frank-Oseen free energy functional, the bulk energetic cost of deformations in a chiral nematic LC reads [23,34]

$$F_{\text{bulk}}^{\text{FO}} = \int d^3 r \left\{ \frac{K_{11}}{2} (\nabla \cdot \mathbf{n})^2 + \frac{K_{22}}{2} \left[ \mathbf{n} \cdot (\nabla \times \mathbf{n}) + \frac{2\pi}{p} \right]^2 + \frac{K_{33}}{2} [\mathbf{n} \times (\nabla \times \mathbf{n})]^2 - \frac{K_{24}}{2} \{ \nabla \cdot [\mathbf{n}(\nabla \cdot \mathbf{n}) + \mathbf{n} \times (\nabla \times \mathbf{n})] \} \right\}, \quad (1)$$

where $p$ is the helicoidal pitch and $K_{11}$, $K_{22}$, $K_{33}$ and $K_{24}$ are the Frank elastic constants describing the energetic costs of splay, twist, bend and saddle-splay deformations, respectively. When the surface anchoring energy is considered finite and taken into account, Eq. (1) is supplemented by the Rapini–Papoular surface anchoring potential [34]

$$F_{\text{surface}}^{\text{FO}} = -\int d^2 r \frac{W}{2} (\mathbf{n} \cdot \mathbf{n}_0)^2, \quad (2)$$

where $W$ is the surface anchoring strength and $\mathbf{n}_0$ is the easy axis orientation (here chosen to be along $\hat{z}$, i.e., along the surface normal of the substrates). The effective strength of surface anchoring relative to bulk elasticity can be further derived by combining Eqs. (1) and (2), and adopting dimensionless units

$$\frac{F_{\text{total}}^{\text{FO}}}{Kp} = \int d^3\tilde{r} \left\{ \frac{\widetilde{K}_{11}}{2}(\widetilde{\nabla}\cdot\boldsymbol{n})^2 + \frac{\widetilde{K}_{22}}{2}[\boldsymbol{n}\cdot(\widetilde{\nabla}\times\boldsymbol{n})]^2 + \frac{\widetilde{K}_{33}}{2}[\boldsymbol{n}\times(\widetilde{\nabla}\times\boldsymbol{n})]^2 \right.$$

$$\left. -\frac{\widetilde{K}_{24}}{2}\{\widetilde{\nabla}\cdot[\boldsymbol{n}(\widetilde{\nabla}\cdot\boldsymbol{n})+\boldsymbol{n}\times(\widetilde{\nabla}\times\boldsymbol{n})]\} + 2\pi\widetilde{K}_{22}\boldsymbol{n}\cdot(\widetilde{\nabla}\times\boldsymbol{n}) \right\}$$

$$-\frac{Wp}{K}\int d^2\tilde{r}\,\frac{1}{2}(\boldsymbol{n}\cdot\boldsymbol{n}_0)^2 \quad (3)$$

where $\tilde{r} = r/p$, $\widetilde{\nabla}= p\nabla$ and $\widetilde{K}_{11}, \widetilde{K}_{22}, \widetilde{K}_{33}$ and $\widetilde{K}_{24}$ are reduced elastic constants scaled by the average elastic constant $K \equiv (K_{11}+K_{22}+K_{33})/3$. Equation (3) can be further simplified by adopting the one-constant approximation by setting $K_{11} = K_{22} = K_{33} = K_{24} = K$

$$\frac{F_{\text{total,1-const}}^{\text{FO}}}{Kp} = \int d^3\tilde{r}\left\{\frac{1}{2}(\widetilde{\nabla}\boldsymbol{n})^2 + 2\pi\boldsymbol{n}\cdot(\widetilde{\nabla}\times\boldsymbol{n})\right\} - \frac{Wp}{K}\int d^2\tilde{r}\,\frac{1}{2}(\boldsymbol{n}\cdot\boldsymbol{n}_0)^2. \quad (4)$$

In this work, the Frank-Oseen free energy minimization is done for material parameters of commonly used commercially available nematics (Table 1) and those under one-constant approximation. We also consider the saddle-splay elastic energy and discuss its effect on the stability of solitons. The surface term in Eq. (3) shows the effective surface anchoring strength is $Wp/K$, the inverse of the extrapolation length $\xi(\equiv K/W)$ normalized by $p$ [34].

To account for disorder and half-integer singular defects in LC director fields with head-tail symmetry, a tensorial theory with a second-rank $Q$-tensor $Q_{ij}$ as the order parameter is required. In Landau-de Gennes modeling, $Q_{ij}$ is a traceless and symmetric tensor defined as [35,36]

$$Q_{ij} = \frac{S}{2}(3n_i n_j - \delta_{ij}) \quad (5)$$

where $S \in [-\frac{1}{2}, 1]$ is the uniaxial scalar order parameter quantifying the molecular orientation fluctuations, $n_i$'s are the $x, y, z$ components of director $\boldsymbol{n}$, and $\delta_{ij}$ is the Kronecker delta. The bulk free energy is given by [35,36]

$$F_{\text{bulk}}^{\text{LdG}} = \int d^3r \left\{ \frac{1}{2} A \text{Tr}(Q^2) + \frac{1}{3} B \text{Tr}(Q^3) + \frac{1}{4} C [\text{Tr}(Q^2)]^2 \right\}$$

$$+ \int d^3r \left\{ \frac{L}{2} \frac{\partial Q_{ij}}{\partial x_k} \frac{\partial Q_{ij}}{\partial x_k} + 2q_0 L \varepsilon_{ikl} Q_{ij} \frac{\partial Q_{lj}}{\partial x_k} \right\} \quad (6)$$

where $A, B,$ and $C$ are material parameters with $A$ depending linearly on temperature $T$, while $B$ and $C$ are effectively temperature-independent material constants. Repeated indices in Eq. (6) are implicitly summed over. The first integral term is the thermotropic free energy and determines the equilibrium scalar order parameter in the bulk $S_{\text{eq}} = \frac{1}{2}\left(\frac{-B}{3C} + \sqrt{\left(\frac{B}{3C}\right)^2 - \frac{8A}{3C}}\right)$. The second integral term is the elastic free energy under one-constant approximation of the long-range director distortions of a chiral LC. $L$ is the elastic constant under one-constant approximation, which can be related to the Frank elastic constant as $L = 2K/9S_{\text{eq}}^2$ and $q_0 = 2\pi/p$. The relative magnitude of thermotropic and elastic energy terms introduces a characteristic length scale for the variation of the scalar order parameter, referred to as the nematic correlation length

$$\xi_N = \sqrt{\frac{L}{A + BS_{\text{eq}} + \frac{9}{2} CS_{\text{eq}}^2}}, \quad (7)$$

which is approximately the size of defects.

The surface energy of uniform anchoring in Landau-de Gennes free energy can be written as [36]

$$F_{\text{surf}}^{\text{LdG}} = \frac{W}{2} \int d^2r \left(Q_{ij} - Q_{ij}^0\right)^2 \quad (8)$$

where $W$ is the surface anchoring strength and $Q_{ij}^0$ is the order parameter tensor preferred by the surface. Homeotropic (perpendicular) anchoring boundary conditions, which are utilized in our experiments, at the confining surfaces are modeled by setting $Q_{ij}^0 = \frac{S_{\text{eq}}}{2}(3e_i e_j - \delta_{ij})$, where $\boldsymbol{e}$ is the surface normal. As with the case in Frank-Oseen free energy, a scaling transformation by $p$ allows the effective anchoring strength to be derived as $Wp/L$.

## *II.2. Energy Minimization*

For solitons to be observed in LCs, they need to be metastable or stable and correspond to local or global minima of the total free energy, including both the bulk and surface terms. Energy minimization is performed by a variational-method-based relaxation routine which includes the boundaries [20,23,24,30]. To minimize the free energy, the field configuration $\boldsymbol{n}(\boldsymbol{r})$ (or $Q(\boldsymbol{r})$) is updated at each iteration based on an update formula derived from the Euler-Lagrange equation of the system. For example, in Frank-Oseen free energy minimization

$$n_i^{\text{new}} = n_i^{\text{old}} - \frac{\mu}{2}[F_{\text{total}}]_{n_i} \quad (9)$$

where the subscript $i$ denotes spatial coordinates, $[F_{\text{total}}]_{n_i}$ denotes the functional derivative of the total free energy functional with respect to $n_i$. The functional derivatives for bulk nodes and boundary nodes are

$$[F_{\text{total}}]_{n_i}^{\text{bulk}} = \frac{\partial f}{\partial n_i} - \nabla \cdot \frac{\partial f}{\partial \nabla n_i} \quad \text{(bulk)} \quad (10-1)$$

$$[F_{\text{total}}]_{n_i}^{\text{bd}} = \frac{\partial f}{\partial \nabla n_i} \cdot \boldsymbol{e} - \frac{Wp}{K}(\boldsymbol{n} \cdot \boldsymbol{n}_0)n_{0i} \quad \text{(boundary)} \quad (10-2)$$

where $\boldsymbol{e}$ is the unit surface normal on the boundary and $f$ is the free energy density in the bulk. $\mu$ is the maximum stable time step in the minimization routine, determined by material parameters and the spacing of the computational grid. $\mu_{\text{bulk}} = \min(d\tilde{x}, d\tilde{y}, d\tilde{z})^2 / 2\max(\widetilde{K}_{11}, \widetilde{K}_{22}, \widetilde{K}_{33}, \widetilde{K}_{24})$ for bulk nodes and $\mu_{\text{bd}} = \min(d\tilde{x}, d\tilde{y}, d\tilde{z}) / 2\max(\widetilde{K}_{11}, \widetilde{K}_{22}, \widetilde{K}_{33}, \widetilde{K}_{24})$ for boundary nodes [20]. The unit modulus of $\boldsymbol{n}(\boldsymbol{r})$ is assured by normalization after each iteration. The steady-state stopping condition is indicated by the change in the spatially averaged functional derivatives over iterations. When this change approaches zero, the system is implied to be in an energy minimum. The Landau-de Gennes free energy is minimized based on the same approach. Six independent components of $Q$-tensor are adopted for improved numerical stability and the traceless condition of $Q$ is then ensured after each iteration [36].

The Frank elastic constants adopted in the energy minimization are based on those independently measured for the nematic hosts (Table 1) [23,24,37]. The 3D spatial discretization of the computational volume is performed on large 3D square-periodic grids with 32 grid points per one pitch $p$ distance equivalent. In Landau-de Gennes modeling, the material parameters are $L = 4 \times 10^{-11}$ N, $A = -0.172 \times 10^6$ Jm$^{-3}$, $B = -2.12 \times 10^6$ Jm$^{-3}$, and $C = 1.73 \times 10^6$ Jm$^{-3}$, corresponding to $\xi_N = 6.63$ nm and $S_{eq} = 0.533$ [36]. The grid length is 10 nm. In both Frank-Oseen and Landau-de Gennes modeling, the spatial derivatives are calculated by finite difference methods with second-order accuracies, allowing the minimization of discretization-related artifacts. Periodic boundary conditions are imposed in lateral directions for simulating both individual solitons and soliton crystals. The lateral dimensions were assured to be sufficiently large for computer-simulating individual solitons such that distortions fully relax to the far-field background on the periphery.

### III. MATERIALS AND METHODS

*III.1. Materials and sample preparation*

LC samples were prepared between two glass substrates, treated with 2 wt% aqueous solution of N,N-dimethyl-N-octadecyl-3-aminopropyltrimethoxysilyl chloride to set homeotropic boundary conditions on the surfaces with an estimated anchoring strength $W \approx 10^{-4}$ Jm$^{-2}$ [38]. To explore the relative importance of surface anchoring effects, the substrates were assembled into thin wedge cells with no spacers on one side (forming an optical contact to ensure vanishing thickness) and 2 μm spherical spacers on the other side. A chiral nematic LC was filled into the cells. The chiral nematic LCs were prepared by mixing a chiral dopant α, α, α, α-tetraaryl-1,3-dioxolane-4,5-dimethanol (TADDOL, Sigma-Aldrich) with a commercially available nematic mixture E7 (Slichem, China) at 2% weight fraction. The resulting chiral LC has a cholesteric pitch $p \approx 1$ μm as measured in a Grandjean–Cano wedge cell [39]. Ferromagnetic LCs are prepared by the same procedures, but additionally utilize small volume fractions of barium hexaferrite nanoplates, which are dispersed in the LC mixtures [24]. Solitonic structures form spontaneously in the cell upon heating to the isotropic phase and quenching back to the nematic phase.

*III.2. 3D nonlinear optical imaging*

The nonlinear optical imaging of $\boldsymbol{n}(\boldsymbol{r})$ of solitonic field configurations was performed using a three-photon excitation fluorescence polarizing microscopy (3PEF-PM) set-up built around an IX-81 Olympus inverted microscope [40,41]. E7 LC molecules were excited via three-photon absorption by using a Ti-Sapphire oscillator (Chameleon Ultra II; Coherent) operating at 870 nm with 140-fs pulses at a repetition rate of 80 MHz [37]. The fluorescence signal was epi-detected by using a 417/60-nm bandpass filter and a photomultiplier tube (H5784-20, purchased from Hamamatsu). An oil-immersion 100X objective with NA = 1.4 was used. The circular polarization state of the excitation beam was achieved using a polarizer and a rotatable quarter-wave retardation plate. No polarizers were utilized within the detection channel. The 3PEF-PM involves a third-order nonlinear process and its intensity scales as $\cos^6 \beta$, where $\beta$ is the angle between the dipole moment of the fluorescent molecule, orienting along $\boldsymbol{n}(\boldsymbol{r})$, and the polarization direction of the excitation beam [23,37]. The 3D 3PEF-PM images for different polarizations of the excitation light were obtained by scanning the excitation beam through the sample volume and recording the fluorescence intensity as a function of 3D scanning coordinates. The images were then post-processed by background subtraction and contrast enhancement. Imaging artifacts in birefringent materials such as beam defocusing and polarization changes are minimized due to the small thickness and pitch of the imaged sample (~1 µm), corresponding to a Mauguin parameter $\frac{p\Delta n}{2\lambda} \approx 0.1 \ll 1$ (Table 1), where $\Delta n$ is the birefringence of the LC and $\lambda$ is the wavelength of the excitation light, whereas the sample thickness still allows for enough axial resolution to resolve the 3D field configurations [40]. Computer simulations of the 3PEF-PM images were based on the $\propto \cos^6 \beta$ dependence of the fluorescence image intensity on the molecular director orientation relative to the polarization direction of the excitation light, quantified by the angle $\beta$.

**IV. Results**

*IV.1. Topology of torons and skyrmions*

We focus on fully non-singular solitonic field configurations of skyrmions and also the solitons accompanied by singular point defects (called "Bloch points" in magnetic structures of chiral bobbers [29] and torons [27,28]). For unit vector fields, 2D baby skyrmions (skyrmions or 2D skyrmions hereafter) are the topological solitons labeled by elements of the second homotopy group, $\pi_2(\mathbb{S}^2) = \mathbb{Z}$ ($\pi_2(\mathbb{S}^2/\mathbb{Z}_2) = \mathbb{Z}$ for nonpolar director fields). The topological degree of a baby skyrmion is the skyrmion number that counts the integer number of times the field configuration wraps around its target space $\mathbb{S}^2$, $N_{\text{sk}} = \frac{1}{4\pi} \int dxdy \, \boldsymbol{n} \cdot (\partial_x \boldsymbol{n} \times \partial_y \boldsymbol{n})$. Figure 1(a) shows full skyrmions that cover $\mathbb{S}^2$ exactly once. An elementary toron is such a baby skyrmion capped by two point defects at its ends, with their existence mandated by the transformation of topology from a skyrmion into the uniform trivial state near the confining substrates [Figs. 1(b) and 1(c)] (note that much more complex types of torons also exist [16,20,23], but they are outside of the scope of this work). Technically, elementary torons can be also referred to as skyrmions, though these are the fragments of skyrmions terminating on singular point defects characterized by the same homotopy group, $\pi_2(\mathbb{S}^2) = \mathbb{Z}$. Such fragments of skyrmions are common in both LCs and magnets, where termination of 2D skyrmions on point defects can be (but is not necessarily) related to confinement [16,20,21,23,27–29,42]. The $\pi_2(\mathbb{S}^2) = \mathbb{Z}$ point defects effectively embed the $\pi_2(\mathbb{S}^2) = \mathbb{Z}$ skyrmions in a localized region of 3D space, but their topology is that of 2D topological solitons terminated on 3D point defects (Fig. 1(c)), different from 3D topological solitons like the $\pi_3(\mathbb{S}^2) = \mathbb{Z}$ (or $\pi_3(\mathbb{S}^2/\mathbb{Z}_2) = \mathbb{Z}$ for nonpolar director fields) hopfions recently studied in condensed matter systems [15,23,24,30,31] and the $\pi_3(\mathbb{S}^3) = \mathbb{Z}$ 3D skyrmions in nuclear and high energy physics [1–4].

In LCs, the order-parameter space (target space) of the nonpolar director field $\boldsymbol{n}(\boldsymbol{r})$ with head-tail symmetry is $\mathbb{S}^2/\mathbb{Z}_2$ (or, equivalently, $\mathbb{R}P^2$), making LC skyrmions elements of $\pi_2(\mathbb{S}^2/\mathbb{Z}_2) = \mathbb{Z}$. However, LC directors can be smoothly decorated with vectors when the directors are orientable, with the orientability determined by the topology of the configuration space and defects in the field [43]. Experimentally, this vectorizing process physically can correspond to dispersing polar nanoparticles, e.g. ferromagnetic nanoplates, in the LC host [Fig. 2(a)] within a preparation procedure that assures their polar ordering and selection of one of the two anti-parallel directions of the nonpolar director field [13,14,24]. Considering this, the orientable director fields are treated as unit vector fields in this work.

### IV.2. Anchoring-controlled stability of torons and skyrmions: numerical analysis

In strongly confined chiral LCs with homeotropic boundary conditions at a sample thickness $d \sim p$, elementary torons are commonly observed particle-like solitons [Fig. 2(b)] [20,21]. Their equilibrium field configurations contain two point defects close to the top and bottom substrates, based on Frank-Oseen free energy minimization [Fig. 2 (c)-(e)]. 2D skyrmions with translationally invariant $\pi_2(\mathbb{S}^2)$ ($\pi_2(\mathbb{S}^2/\mathbb{Z}_2)$) field topology along the symmetry axis were also observed when the surface anchoring is softened [16] or when the size of the overall structure is reduced [44], though possible inter-transformations between elementary torons and skyrmions as a result of confinement and boundary conditions have not been systematically considered. We investigate numerically the dependence of structural stability of solitonic structures on thickness to pitch ratio $(d/p)$ and effective anchoring strength characterized by a dimensionless quantity $Wp/K = (\xi/p)^{-1}$ based on various material parameters, including those under one-constant approximation and those of commonly used nematic LCs [Fig. 3 (a)-(d) and Table 1]. The initial state of energy minimization is a toron, which mimics the experimental condition that such twisted localized configurations form in the bulk upon quenching of LCs from isotropic to nematic phase. As shown in Fig. 3(a)-(d), torons are stable with respect to translationally invariant skyrmions or other states when the surface anchoring is sufficiently strong and the thickness sufficiently thick. At a smaller thickness, torons' stability is lost with respect to the uniformly unwound state, with the boundary between torons and uniform states also dependent on the anchoring strength. 2D skyrmions are stable for weak anchoring. The transition between torons and skyrmions happens for $10 \lesssim Wp/K \lesssim 30$, depending on material parameters and also on sample thickness. Practically, the transformation of torons to skyrmions can occur at cell thickness of tens of micrometers for soft boundary conditions and only for sub-micrometer-tick cells in case of strong boundary conditions. Stable energy-minimizing field configurations of 2D arrays of torons and skyrmions are also stabilized, as shown in Figs. 3 (e)-(l), sharing the same stability conditions as their individual localized counterparts. The energies of the solitonic states are compared to that of the uniform state, where a yellow (cyan) background in Figs. 3(a)-(d) indicates the solitons have lower (higher) energies than the uniform unwound state. At a larger thickness and weaker surface anchoring strength, the translationally invariant configuration (TIC) [45] is another topologically trivial background state that competes with the uniform state and can be of lower energy. Regions with the uniform state or TIC being the lower energy state are labeled with filled or unfilled squares, respectively. TIC can also host solitons such as skyrmions or torons [46], though a detailed study of such structural configurations is beyond the scope of this paper. Figures 3(a)-(d) show that a solitonic 2D skyrmion or toron is the lowest energy state within certain parameter regions, out of the non-exhaustive competing states considered in this work. An exhaustive ground-state stability diagram analysis could encompass all trivial and topological states, including the modulated TIC [45], various cholesteric finger textures [45], complex torons and hopfions [23], etc., but this is also beyond the scope of our present work.

The stability of torons, skyrmions and the uniform state can be understood by their structures and the frustration between different terms in free energy. For example, the double twist skyrmionic structure within torons lowers the system's energy associated with twist deformations as compared to the uniform unwound state. However, the hyperbolic point defects within torons involve large elastic distortions and even localized regions of opposite-handedness twist to that of the molecular chirality of the bulk LC [22], making them energetically costly. In the strong anchoring regime, the existence of singular defects is a result of the presence of strong boundary conditions at the substrates, which compels the field configuration to transform from a 2D skyrmion in the horizontal midplane and within the sample bulk to the uniform state near the two confining substrates. This transformation is accomplished by a singular defect, which also shows a complete coverage of the order-parameter space and a nontrivial $\pi_2(\mathbb{S}^2)$ ($\pi_2(\mathbb{S}^2/\mathbb{Z}_2)$) field topology on the enclosing surface [Fig. 1(c)]. This structure is a vivid demonstration that topologically distinct structures of the skyrmion tube within the LC bulk and the uniform state at the cell substrates can be inter-transformed only with the help of other topologically nontrivial configurations, which in this case are point defects. In the weak anchoring regime, the system lowers energy by pushing the point defects away from the bulk, and eventually beyond the substrates, making them virtual. A toron thus becomes a 2D skyrmion, with its field structure translationally invariant along its symmetry axis along $\hat{z}$. A sequence of snapshot of this transformation under weak anchoring are shown in Fig. 4. Note that surface defects called "boojum" can form as an intermediate state or at suitable bulk and surface energy interplay conditions [47,48], as shown in Fig. 4(b). In the strong anchoring regime and with small sample thickness, the surface energy dominates and aligns the field along the easy axis, resulting the structures to unwind and relax to the uniform state. This also can be understood because twisted structures with twist rate given by the equilibrium pitch cannot be embedded at low energetic costs within a sample with strong boundary conditions when the pitch is significantly larger than cell thickness, as a result of the surface-bulk energetic competition at given $\xi$ and $d$.

The elastic constant anisotropy has a noticeable effect on the stability of different solitons, albeit the dependence of stability on thickness and effective anchoring follow the same general trend (Fig. 3). In particular, the transition between the uniform state and the toron happens at different thicknesses for material parameters of 5CB, E7, and a one-constant approximated system (Fig. 3). The thicknesses at which the uniform state loses stability to TIC ($d/p = 1.45$ for 5CB, $d/p = 1.2$ for E7, and $d/p = 0.5$ for a one-constant-approximated LC system), are also markedly different. This is due to splay and bend deformations in a one-constant approximated LC system being less penalized as compared to the cases of E7 and 5CB (Table 1) with the twist elastic constant significantly lower than the bend and splay constants. Thus, $n(r)$ is more likely to adopt splay and bend deformations within the one-constant-approximation at a smaller thickness in a confined geometry, favorable for the stability of torons and TIC. The inclusion of the saddle-splay elastic energy in the free energy functional has little effect on stability in the case of E7 when taking $K_{24} = K_{22}$ [Fig. 3(d)]. As a total divergence, the saddle-splay energy can be considered as a surface elasticity and renormalizes the surface anchoring strength within the finite-anchoring-strength modeling considered here [49], though the saddle-splay term also contributes to the energetics of point defects within torons.

### *IV.3. Anchoring-controlled stability of torons and skyrmions: experiments*

The competition between surface anchoring and elasticity within the studied confined LCs depends on the magnitude of the average elastic constant $K$, surface anchoring coefficient $W$, $d$ and $p$. The boundary conditions at confining surfaces are violated for weak $W$ and/or for thin cells, when surface anchoring extrapolation length $K/W$ becomes of the same order of magnitude or even larger than $d$ and $p$, facilitating appearance of translationally invariant skyrmions. In the opposite regime, for $K/W$ much smaller than $d$ and $p$, torons with point defects are expected to appear. To observe experimentally the structural transition between torons, 2D skyrmions, and the uniform state, a confined chiral LC is prepared such that the parameter $Wp/K$ is at the border between weak and strong anchoring regimes. While systematic control of $W$ using a single type of surface functionalization is difficult, the effective role of the surface anchoring strength can be weakened by reducing $p$ and $d$ while keeping $W$ and $K/W$ fixed. Note that the choice of the LC material also determines the interplay between surface and bulk energies due to different average elastic constants $K$. Listed in Table 1 are the elastic constants of a few commercially available nematics, where $K$ can differ by up to a factor of 2. For the nematic LC E7, $K \approx 10$ pN (Table 1) and $W \sim 10^{-4}$ Jm$^{-2}$ (strong homeotropic conditions set in the experiment), and the effective threshold between strong and weak confinement regimes is $Wp/K \approx 20$ [Fig. 3(c)]. This corresponds to the characteristic pitch $p \approx 1$ μm and thickness $d \approx 1$ μm at which toron-skyrmion inter-transformations and coexistence are expected to take place. Under such conditions, the competition between the thickness $d$ and extrapolation length $\xi$ allows toron-skyrmion inter-transformations and coexistence to take place by varying the thickness $d$ [Fig. 3(c)].

To validate the above predictions experimentally, a chiral LC based on E7 with $p \approx 1$ μm was filled into a wedge cell, where the thickness varies linearly from zero to 2 μm. In the cell region where $d \approx 1$ μm, torons, 2D skyrmions and even hybrid structures of torons and skyrmions were observed (Figs. 5-7). The distribution of torons and skyrmions were correlated with the cell thickness. In regions with $d \gtrsim 1$ μm, torons prevail and 2D translationally invariant skyrmions were rarely observed. Cholesteric fingers of the first type [45] start to grow as $d$ further increases to be substantially larger than 1 μm. As $d$ decreases, torons disappear and 2D skyrmions emerge instead. This is consistent with the thickness dependence of structural stability between torons and 2D skyrmions near the anchoring threshold for the same material shown in Fig. 3(c). At even smaller thickness, $\boldsymbol{n}(\boldsymbol{r})$ becomes completely unwound and homeotropic. This is because of the initial state forming after thermal quenching being thickness-dependent, favoring the unwound state at smaller thicknesses, as discussed above.

The polarizing optical micrographs and 3D nonlinear optical images reveal the detailed structural difference between torons, 2D skyrmions and hybrid structures (Figs. 5-7). Torons contain the two point defects near the substrates (Fig. 6), which are revealed by the 3D nonlinear optical imaging (Fig. 6(e),(f)), whereas a 2D skyrmion is translationally invariant throughout the sample thickness (Fig. 5). The polarizing optical micrographs also show distinct patterns that allow for distinguishing between a toron and a 2D skyrmion when co-existing under the same sample conditions on the basis of the accumulated phase retardation resulting from their respective 3D director distortions. We note that though sharing the same field structures and materials, the polarizing optical micrographs of torons in Fig. 6a and Fig. 2b are distinctly different because of the corresponding disparity in their dimensions and cell thicknesses, where polarized light interreference and different phase retardations and rotations of the long axes of polarization ellipse while light is traversing the sample result in very different colors and appearance. This shows that the underlying 3D LC structures may not be sufficiently inferred from the optical polarizing micrograph textures, when certain material or sample parameters such as optical birefringence, elastic anisotropy, sample thickness, etc., are different. In experiments, local variation in the coated alignment layer can cause the anchoring strength to be slightly different, asymmetric on the two substrates, along with thermal fluctuations, resulting in only one of the point defects escaping. This gives origins to the toron/skyrmion hybrid structure shown in Fig. 7. Such toron/skyrmion hybrid structures are also observed in solid-state chiral magnets where they are referred to as "bobbers", stabilized by an applied magnetic field in the conical state [29]. Importantly, the relative stability of torons, skyrmions and their hybrids corelates with the numerical stability diagrams for different studied materials (Fig. 3). While the experiments discussed above focused on relatively thin cells and rather strong perpendicular anchoring of $W \sim 10^{-4}$ Jm$^{-2}$, much softer perpendicular boundary conditions are often also used in experiments with $W \sim 10^{-5} - 10^{-6}$ Jm$^{-2}$, often allowing for the realization of translationally invariant 2D skyrmion and skyrmion bag structures in cells of larger thickness [16,50,51], consistent with the structural stability diagrams presented in Fig. 3.

### *IV.4. From two- to one-dimensional topological solitons: twisted walls versus fingers*

The existence of singular defects capping the ends of a 2D topological soliton (the baby skyrmion) within the elementary toron is a result of stitching together distinct topological states of skyrmion and the uniform state. Equally, this also is the result of geometrical frustration between bulk (favoring the twisted skyrmion) and surface (favoring uniform state) free energy terms when a lower-dimensional 2D soliton is embedded in a higher-dimensional 3D space with strong boundary conditions. In particular, when a 2D skyrmion is embedded in the 3D space, strong perpendicular boundary conditions require uniform states at the surfaces. Therefore, the skyrmion terminates at singular defects which serve the purpose of mediating the transformation between topologically distinct bulk and surface states, forming a toron. The point defects are classified by the same $\pi_2(\mathbb{S}^2)$ ($\pi_2(\mathbb{S}^2/\mathbb{Z}_2)$) topology as the 2D skyrmion, even though the singular points are very different objects from the baby skyrmions. Could a similar situation arise for co-existence of solitonic and singular topological structures in other dimensions, say for the 1D topological solitons, which require only $\mathbb{R}^1$ to exist, when confined/embedded in $\mathbb{R}^2$ or even in $\mathbb{R}^3$?

In condensed matter systems like LCs and magnets, the solitonic wall structures are rather common [34,44,52,53]. Moreover, twisted walls in these condensed matter systems can be stabilized by the medium's chirality, similar to that of skyrmions studied above [44]. To embed the solitonic twisted walls in the uniform far-field background for unit vector fields, only an integer numbers of $2\pi$-twist of this unit vector field can be smoothly embedded, which is characterized by the $\pi_1(\mathbb{S}^1) = \mathbb{Z}$ homotopy group. The situation is different in nonpolar LCs, where integer numbers of π-twist can be embedded in the nonpolar far-field background of the LC director [Fig. 8(a)]. Within the solitonic LC twisted wall, the nonpolar directors $\boldsymbol{n}(\boldsymbol{r})$ are effectively confined in a 2D plane [Fig. 8(b)] and the order-parameter space is $\mathbb{S}^1/\mathbb{Z}_2$, or, equivalently, the real projective line $\mathbb{R}P^1$. Topologically, $\mathbb{R}P^1$ (or $\mathbb{S}^1/\mathbb{Z}_2$) is homeomorphic to a circle, the 1D sphere $\mathbb{S}^1$ [Fig. 8(c)], and therefore the first homotopy group of $\mathbb{R}P^1$ is effectively the first homotopy group of $\mathbb{S}^1$, $\pi_1(\mathbb{S}^1) = \mathbb{Z}$. A $\pi$ rotation of director within the elementary solitonic LC twisted wall covers the $\mathbb{S}^1/\mathbb{Z}_2$ order parameter space exactly once, also smoothly embedding into the uniform far-field director background, which is different from the case of a vectorized field (Fig. 8(d)). While the structure of such solitonic twisted wall implies translational invariance in the plane of the wall, how could it be embedded in a finite-size sample in $\mathbb{R}^2$ or in $\mathbb{R}^3$ when the boundary conditions within the LC sample impose a uniform configuration at confining surfaces? Below we show how this difference between the topologically distinct twisted wall state in the LC bulk and the uniform state at LC surfaces leads to the emergence of singular twisted lines and formation of the so-called cholesteric finger structure.

A 2D cross-section of an LC cholesteric finger of the third kind, the CF-3 [16,45,54], can be hypothetically considered as a twisted wall, a 1D $\pi_1(\mathbb{S}^1)$ topological soliton in $\mathbb{R}^1$, terminating at two $\pi_1(\mathbb{S}^1)$ half-integer defects near the substrates with strong perpendicular-anchoring boundary conditions for director orientations confined in the yz-plane within an effectively 2D nematic [Fig. 8(e), top] (note that $\pi_1(\mathbb{S}^2/\mathbb{Z}_2)$ describes both the twisted wall and the disclinations for fully 3D director orientations [Fig. 8(e), bottom]). Figure 8(e) shows the computer-simulated cross-section of a CF-3 normal to the direction of translational invariance, based on energy-minimization of the Landau-de Gennes free energy. The two localized regions with reduced scalar order parameter within the finger's cross-section correspond to two singular defect cores of twist disclinations near the substrates. Note that unlike the twisted wall [Fig. 8(d)], the director field of the two twist disclinations and the entire CF-3 is non-orientable and vectorizing it creates extra discontinuities in the form of additional wall-like defects. The defects extend along the translationally invariant direction and form defect lines. The nonpolar director field along a loop around the half-integer defect lines form Möbius strips tangent to the director, forbidding a smooth vectorization of the field [Fig. 8(f)]. Under weak anchoring conditions, the defect lines escape beyond the surfaces, rendering the field configuration topologically invariant across the thickness, analogous to the case of toron/skyrmion transformation. The structural stability between CF-3, twisted wall, and the uniform state depending on the sample thickness and the effective anchoring strength is shown in Fig. 8(g). As with torons, CF-3s are stable at large enough thickness and anchoring strength, though the thickness required for stable CF-3s is about half of that required for stable torons, according to our modeling under one-constant approximation. In the weak anchoring regime, deviation of director orientation away from the anchoring direction at surfaces is allowed and twisted walls become the structurally stable configuration. The numerically derived stability of CF-3 and twisted wall is consistent with experimental observations [Fig. 8(h)-(l)], where CF-3s are observed in common LC nematics (Table 1) with $p \sim 10$ μm and $W \sim 10^{-4}$ Jm$^{-2}$ [16,45,54], corresponding to strong effective anchoring in Fig. 8(g), while 1D twisted walls are ubiquitously observed when $p \lesssim 1$μm [44]. The cross-sectional 3PEF-PM images of a twisted wall and a CF-3 reveal their detailed structural difference, where the two localized half-integer defect lines are evident within a CF-3 finger, whereas the twisd wall remains largely translationally invariant across the sample thickness [Fig. 8 (i)-(k)]. The absence of high-energy defect lines in twisted walls also allow them to be more abundant within the sample, lowering the overall elastic energy through twisting, while a CF-3 has a tendency to shrink due to the defect-induced tension [Fig. 8

(h) and (l)]. Both twisted walls and cholesteric fingers can co-exist at intermediate ranges of pitch and thickness within 1-10 μm. The twisted walls looped into circles can form skyrmions and skyrmion bags [50] (Fig. 9), whereas closed loops of CF-3s can inter-transform into torons. Because the exact stability requirements for all these structures are slightly different, anchoring, LC material parameters, confinement, temperature, fields, various dopants and other means can be used to drive inter-transformations between these spatially localized director configurations. On the other hand, our numerical analysis-based findings also imply that, depending on strong or soft surface anchoring boundary conditions, one can stabilize twisted walls and skyrmion bags (Fig. 9) or, instead, CF-3s and their closed loops in cells of thickness ranging from hundreds of nanometers to tens of micrometers, consistent with experiments [50]. While above we considered mainly just the concept of embedding the twisted walls within 2D cross-sections of translationally invariant CF-3 finger structures, which in real experimental samples often span the entire lateral extent of the LC cells [Fig. 8(l)], these fingers can also terminate within the sample with the uniform unwound far-field background, which corresponds to embedding such twisted solitonic walls in $\mathbb{R}^3$. Within a fragment of the CF-3 in a homeotropic sample, the bottom and top disclinations close into a loop, so that the twisted wall is confined within a uniform background in 3D.

**V. Discussion**

*V.1. Implications to other condensed matter systems and other solitonic structures*

Beyond the elementary torons and skyrmions discussed in this work, various other solitonic structures can be stabilized in confined chiral LCs. Some of these structures are accompanied by point defects, such as twistions, multiple-$\pi$-twist torons [22,23], and cholesteric fingers of the third and fourth types containing singular line defects [45], while others are smooth field configurations, such as hopfions [15,23,24], heliknotons [25], cholesteric fingers of first and second kinds (referred to as CF-1s and CF-2s) [23,55]. Strong boundary conditions not only prompt existence of energetically costly singular defects while embedding topological solitons within samples that have topologically trivial structure, but are also often necessary in stabilizing solitons with localized regions of high elastic energy cost near the substrates. For example, surface confinement was found to be relevant in stabilizing hopfions in both LCs and solid-state chiral magnets, where the boundary condition is fixed by the interfacial perpendicular magnetic anisotropy in the latter case [30]. By engineering the surface interactions, analogues of toron/skyrmion transformation for hopfions and other solitons could also be observed and have been recently considered in magnetic systems [28,30], enriching the variety of stable topological solitons in chiral materials. Our findings demonstrate that additional studies of more complete structural diagrams will need to be in future performed to involve an exhaustive zoo of possible field configurations (say, by comparing stability between skyrmions, torons, fingers, and hopfions, under different surface confinement conditions), though this is beyond the scope of our present work. The important insight from our study for the field of LC solitons, however, is that surface anchoring is an effective knob for tuning stability of all these different solitonic and related structures.

The free-energy models of chiral LCs shares a close resemblance with chiral magnets. The bulk energy in Frank-Oseen functional in Eq. (1) reduces to the micromagnetic Hamiltonian of chiral ferromagnets under one-elastic-constant approximation [50]. Coupling terms in ferromagnets, such as Zeeman coupling and magnetocrystalline anisotropy, which are critical in soliton stability [56,57], also find analogues in LCs. For example, the quadratic dielectric and diamagnetic coupling in LCs is equivalent to the uniaxial magnetocrystalline anisotropy in magnets, with the positive (negative) dielectric anisotropy of the LC molecules corresponding to an easy axis (plane) anisotropy. In ferromagnetic LCs, the magnetic moments of ferromagnetic nanoparticles couple linearly to an externally applied magnetic field, resulting in a Zeeman-like coupling term. In magnets, surface energy terms at interfaces where the inversion symmetry is broken can be significant [58,59], much like the surface anchoring energy in LCs. This resemblance allows the prediction of solitons stabilized in LCs to also be stabilized in magnetic systems, including skyrmions [9,10], torons [27,28], skyrmion/toron hybrids [29], hopfions [28,30,31], etc. On the other hand, the elastic constant anisotropy (absent in magnetic systems) in realistic LC materials provides an extra means of controlling energy frustration and stability of solitons (Fig. 3). Our results show (Fig. 3) that for material parameters of 5CB and E7, skyrmions and torons can be of lower energy than the background uniform state in extended parameter regions. This is consistent with experimental observations where skyrmions or torons form spontaneously after thermal quenching or electrically promoted hydrodynamic instability [21]. For a one-constant approximated LC system, the parameter regions with skyrmion or toron being the lower energy state in a uniform background are much smaller or do not exist (Fig. 3), due to the relatively low energetic cost of forming TIC [Fig. 3(a)]. This suggests that the elastic constant anisotropy has to be taken into account for a realistic modeling of topological soliton stability in LCs. Listed in Table 1 are the elastic constants of a few common and commercially available nematics, for which the parameter ranges of stability vary significantly due to different elastic constant anisotropies and values of elastic constants relative to typical surface anchoring coefficients (Fig. 3). LCs with exceptionally large elastic anisotropy include twist-bend nematics [60], splay-bend nematics [61] and lyotropic chromonics [62], where the ratio between different elastic constants can be disparate by orders of magnitude. In future studies, it will be of interest to explore how such elastic ultra-anisotropies can stabilize or de-stabilize formation of various topological solitons.

While free energy functionals take different forms in different physical systems, ranging from LCs to magnets and high energy models [1,11,17,34,35,56], the Derrick-Hobart theorem is a powerful tool of inferring stability of localized solitons in 2D and higher dimensions based on the energy of the spatially rescaled field configurations [32]. According to the theorem, a free energy with only quadratic spatial derivatives cannot host stable solitons in 2D or 3D [63–65]. Common approaches to evade the Derrick-Hobart theorem and establish soliton stability include incorporating in the free energy higher-order [1] or linear [64,65] spatial derivative terms. In chiral LCs, the molecular chirality introduces a linear chiral term, which sets an intrinsic length scale $p$ and tends to stabilize the hosted topological solitons, which is directly analogous to the Dzyaloshinskii–Moriya interaction in chiral magnets [11,66,67]. In addition to chirality, elastic constant anisotropy and confinement, surface anchoring also provides alternative routes for establishing soliton stability in LCs, effectively making the conclusions of the Derrick-Hobart theorem non-detrimental for the stability of high-dimensional solitons in these systems because the theorem applies only to unconfined physical systems. These findings may provide inspirations for approaches to stabilize high-dimensional topological solitons in other physical systems, as already seen for the case of chiral magnets [30].

*V.2. Vector versus tensor modeling: from topological solitons to merons and blue phases*

In this work, both Frank-Oseen and Landau de-Gennes energy minimization are performed complimentarily. The simpler Frank-Oseen free energy modeling is adequate in the uniaxial nematic regime with orientable field configurations, while the Landau de-Gennes modeling encompasses the full degree of freedom of the order and orientation of a nematic LC and is able to model non-orientable line fields. The elastic constants in Frank-Oseen free energy directly relate to the experimentally measurable moduli of different elastic modes in a nematic LC, while in Landau-de Gennes, expansion up to third order of the $Q$-tensor order parameter in the elastic energy is necessary to adequately lift the degeneracy between the elastic constants [68]. To compare between simulation results from Frank-Oseen and Landau-de Gennes modeling, the elastic constant anisotropy must also be taken into account, in addition to other fundamental differences. For example, Landau-de Gennes free energy models LCs at a temperature near the nematic-isotropic transition, while the Frank-Oseen free energy models a uniaxial nematic at a constant order. By vectorial modeling for orientable field configurations (skyrmions and torons) and tensorial modeling for non-orientable ones (CF-3s), we have gained insights into the conditions of transformations between purely nonsingular topological solitons and such structures terminating on singular defects. Tensorial modeling that accounts for elastic anisotropy will be needed to comprehensively compare stability of all different possible structures for the large parameter spaces of LC materials.

Beyond the various structures discussed above, LC blue phases can emerge as thermodynamically stable states under suitable conditions and typically in a narrow temperature range near the chiral nematic-isotropic transition. The blue phases consist of fractional skyrmion (meron) tubes and singular defect lines typically arranged into crystalline lattices with different symmetries [69]. In confined geometries, 2D lattices of merons (or half-skyrmions) could also emerge in a chiral LC thin film under certain temperature and anchoring conditions [70,71]. LC blue phases have attracted much interest in studying self-assembly of fractional skyrmion tubes and defect lines, and also in applications including fast-switching display modes and other electro-optic and photonic applications [72,73]. On the other hand, a lattice assembled by full skyrmions is fully nonsingular, and could compete in free energy with meron lattices, where a delicate balance between energetic costs of disorder in defects and elastic deformations could be tipped by various system parameters. A systematic study of such comparison between meron and skyrmion lattices in 2D and 3D, considering different temperature, anchoring, elasticity, and externally applied fields, etc., is important and could facilitate the discovery of new topologically nontrivial thermodynamically stable phases. However, while numerical studies of blue phases and various fractional skyrmion lattices have been performed previously, to the best of our knowledge, they did not fully account for elastic constant anisotropy through including higher-order gradient terms in the tensorial modeling. Our findings point out the need of such future investigations.

Although the reasoning above calls for more complex $Q$-tensor based modeling with high-order gradient terms, much simpler types of modeling also have their own important merits as they can translate insights between different branches of condensed matter and beyond. For example, our recent work on skyrmion bags involved effectively 2D vectorial modeling that could be directly connected and compared to such modeling done for magnetic systems [50]. Under what conditions can one justify doing 2D versus 3D modeling? Solitons in this work are modeled by 3D energy minimization. However, under certain conditions, the field configurations of solitons are translationally invariant or have small conformational variations across the thickness, and therefore can be modeled by 2D simulations. The relevant parameters are the normalized thickness and the extrapolation length ($d/p$ and $\xi/p$). As shown in Fig. 3(a)-(d), torons, whose field topology varies across the thickness, only emerge as $d/p$ is large and $\xi/p$ is small, namely when the thickness is large to allow distortion along $\hat{z}$ and the effective anchoring strength is strong to confine the topology-transforming defects. In the weak anchoring regime, solitons such as elementary skyrmions or high-charge skyrmion bags have translationally invariant structure and indeed can be effectively modeled by 2D simulations [Fig. 9(a),(b)]. In 2D simulations, invariance is assumed across the thickness and the surface anchoring term $-\frac{1}{2}W(\boldsymbol{n}-\boldsymbol{n}_0)^2$ translates to an effective bulk anisotropy term $-\Lambda(\boldsymbol{n}-\boldsymbol{n}_0)^2$ with coefficient $\Lambda \equiv W/d$. One can then directly compare the predictions of 2D and 3D modeling under such conditions. Figures 9 (c),(d) show the stable field configurations of solitons in 3D, and their horizontal midplane cross-sections. The midplane cross-sections of field configurations modeled with the 3D energy minimization approach are in an excellent agreement with pure 2D simulations, while variation in field configurations along $\hat{z}$, visualized by preimages, are visible due to the finite thickness. The variation becomes smaller and eventually vanishes as $d/p$ approaches 0, rendering a true 2D system, which is a regime when the 2D modeling is fully adequate (though it can provide great insights even somewhat away from this regime) [50]. While 3D tensorial modeling with higher-order terms is important for a comprehensive analysis of stability of the large zoo of solitonic and singular defect structures that can occur, the 2D simplified modeling is, in turn, important for technology-transferring insights from the studies of LC solitons to understanding related phenomena in other physical systems [50].

## V.3. From fundamental research to applications of topological solitons

The study of LC solitons is of fundamental interest for demonstrating various field theories and the mathematical knot theory, as well as inspirational in understanding solitons in other physical systems. It is also beneficial in understanding the stability/metastability of thermodynamically stable topological phases in condensed matter systems, such as the LC blue phase [69] and skyrmion A-phase in ferromagnets [9,10], and may inspire discovery of other such phases comprising solitons and fractional skyrmions. The experimental observations and analyses of LC solitons benefit from their outstanding accessibility, where laser manipulation and 3D optical imaging techniques can be utilized to decipher their complex structures [40], while 3D real-space imaging in many other physical systems have so far remained challenging. In addition, LC solitons are adaptive and reconfigurable particles, providing a platform for fundamental research involving systems of many particles, such as crystal assembly [21,25,26], active matter [51], artificial spin ice [19], etc.

The energetic landscape of a frustrated system is complex and multistable in nature. As a result, the conditions for (meta)stability could overlap and different solitons could coexist under the same condition in a confined chiral LC [15,16,22,23]. The energetic difference between coexisting solitons can be thousands of $k_\mathrm{B}T$ or higher, while energetic barriers prohibit transition between the multi- and meta-stable solitonic states. These barriers can be topological in nature, where costly singular defect states are mandatory to mediate the transition between topologically distinct solitonic states [15], or they can simply be due to the nonexistence of connecting energetic pathways that involve no topology transformations. Some of these barriers are explained by the concept of contact topology [74]. Surface anchoring and confinement, as demonstrated in this work, can be an effective means of controlling soliton stability. Along with externally applied electric and magnetic fields, which are able to substantially disturb the system's energetic landscape and switch between LC solitonic states [15], the stability of solitons can be controlled and switched by a versatile set of control knobs. Since the hosting LCs are optically birefringent and so are the LC solitons, this may enable potential applications in the forms of soliton-based LC display modes and electro-optic devices. In particular, multi-stability could be potentially a useful feature allowing one to develop bi-stable and multi-stable displays with new functionalities and with energy efficiency better than that of current displays. The topological and particle-like nature and the inter-soliton interaction allow LC solitons to assemble into 1D, 2D and 3D periodic superstructures with electrically tunable spatial periodicities [25,75,76], potentially leading to applications of adaptive and reconfigurable soliton-based diffraction optics. Other applications of LC solitons include optical memory devices, optical logic devices, microfluidic devices, etc. Since the anchoring-based means of controlling these structures are based on that of controllable surface treatments based on the developments of the multi-billion display industry, these methods of controlling solitons are already highly technological, and, thus, also of high applied interest.

VI. Conclusions

To conclude, we have demonstrated the effect of surface anchoring on the structural stability of LC torons, 2D skyrmions, fingers, and twisted walls based on numerical energy minimization of Frank-Oseen and Landau-de Gennes free energy functionals. Our numerical results are supported by experimental observations, based on 3D nonlinear optical imaging of director fields that reveal unambiguously the respective solitonic structures. We found the structural stability of these solitons depend critically on the sample thickness and the effective anchoring strength. The stability of solitons also substantially depends on the elastic constant anisotropy, making the one-constant approximation of elastic constants unsuitable for quantitative analyses. Strong boundary conditions materialized by surface alignment layers mandate a change in topology across the thickness, transforming topological solitons in lower dimensions (twisted walls and 2D skyrmions) to solitons dressed by defects (CF-3s and torons). A wealth of 2D or 3D particle-like solitons can thus be stabilized by engineering the surface anchoring conditions. Moreover, since the effective anchoring strength depends on $p$, which is the characteristic length scale of a solitonic structure, we found the stability of solitons and other field configurations at different length scales can be distinctly different. New types of solitonic structures may be discovered simply by changing the length scale and perturbing the frustration between the bulk and surface energy. The effect of surface energy is not limited to confined LC systems. In other condensed matter systems, for example, the solid-state magnets, surface effects such as the interfacial Dzyaloshinskii–Moriya interaction and the interfacial anisotropy can emerge at the heterojunction between different materials [58,59], which help stabilize various kinds of 2D and 3D solitons [12,30]. In addition to applied fields, confinement and anchoring offer extra control knobs for soliton stability, bringing about potential soliton-based information displays and other technologies. Based on a general framework for ordered materials with interplaying bulk and interfacial effects, our findings could be generalized to the stability of solitons in other material systems.

Acknowledgement

We acknowledge discussions with P. J. Ackerman, M. Tasinkevych, H. Sohn, S. Žumer and funding from the National Science Foundation (NSF) grant DMR-1810513. This work utilized the RMACC Summit supercomputer, which is supported by the NSF (Grants No. ACI-1532235 and No. ACI-1532236), the University of Colorado Boulder and Colorado State University.

**Figures:**

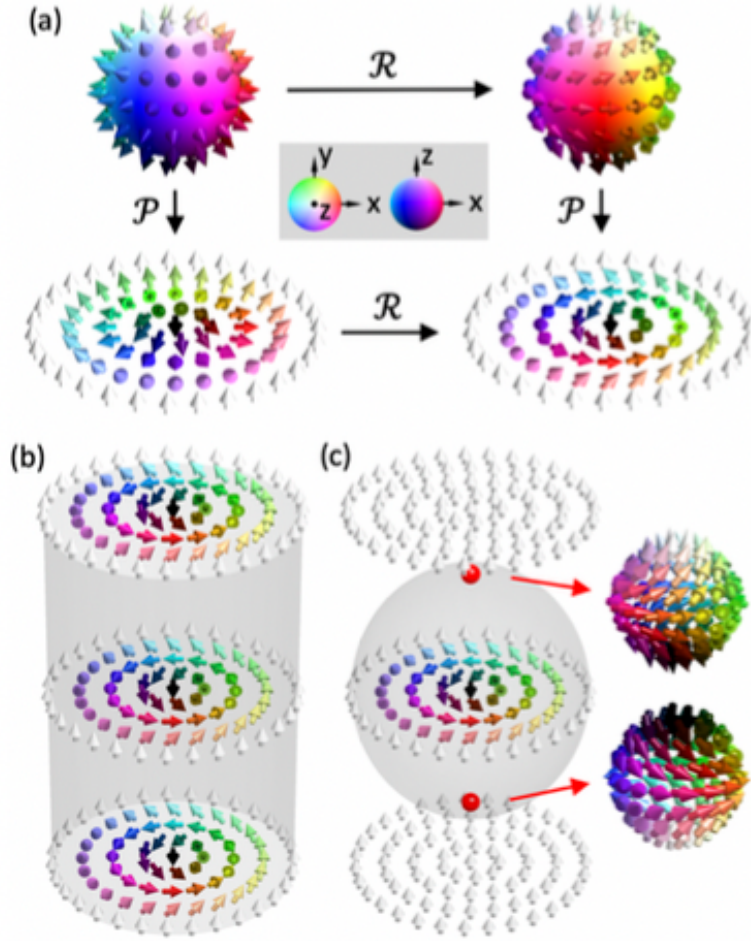

FIG. 1. Skyrmions. (a) Skyrmions in $\mathbb{R}^2$ (bottom) can be mapped bijectively from field configurations in $\mathbb{S}^2$ (top) through stereographic projections ($\mathcal{P}$). The Neel-type (bottom-left) and Bloch-type (bottom-right) skyrmions are related by a rotation ($\mathcal{R}$) of vectors. The vector orientations are shown as arrows colored according to their orientations on target $\mathbb{S}^2$, as shown in the inset. (b),(c) A translationally invariant skyrmion along its symmetry axis and a skyrmion terminating at point defects due to boundary conditions, respectively. The detailed field configurations on spheres around the point defects are shown.

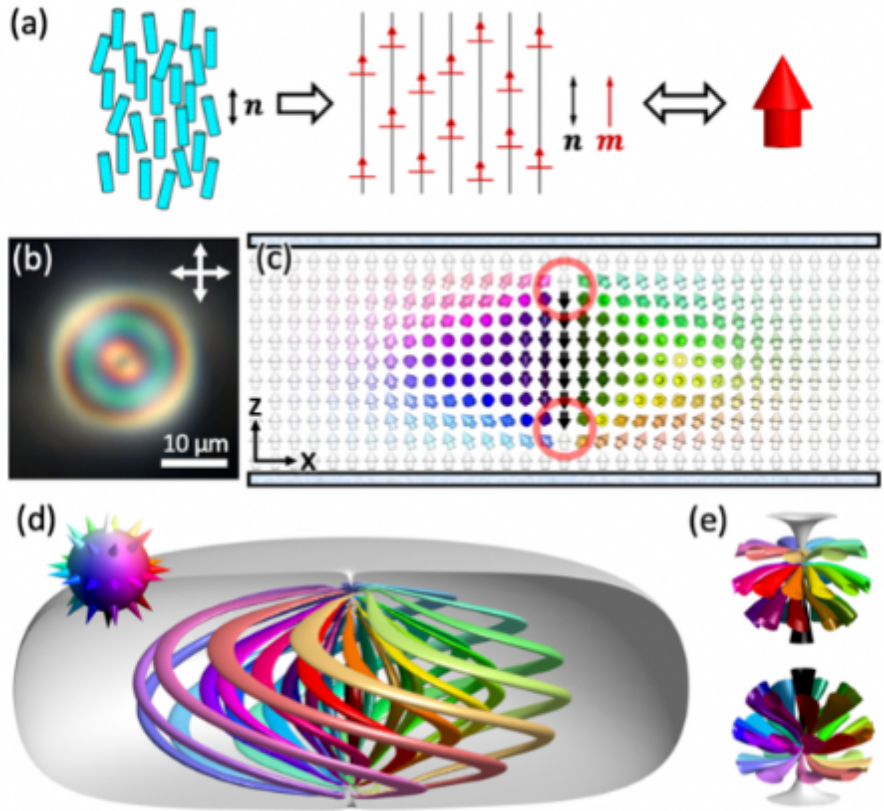

FIG. 2. LC torons. (a) The director $n(r)$ along the average orientation of anisotropic mesogens in nematic LCs can be vectorized by the dispersion of ferromagnetic particles with magnetization vectors $m(r)$ aligning with $n(r)$. The orientation of vectorized $n(r)$ is visualized by a 3D arrow. (b) Polarizing optical micrograph of a LC toron in a cell with thickness $d = 10$ μm. (c) The field configuration of a toron in the vertical mid-plane cross-section with the field shown as arrows colored according to the inset in Fig. 1(a). The point defects are marked by red circles. (d) The preimages of a toron of orientations in target $\mathbb{S}^2$ shown as cones. A preimage is the region in $\mathbb{R}^3$ that maps to a certain orientation on target space $\mathbb{S}^2$. (e) Detailed structures of preimages around the top and bottom point defects of the toron in (d).

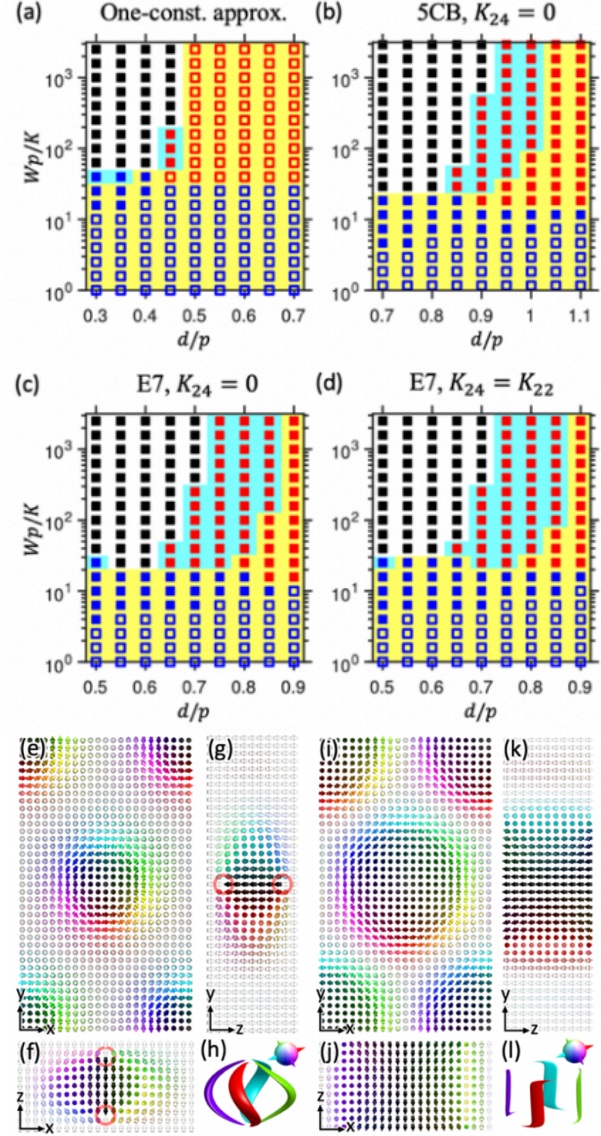

FIG. 3. Structural stability between 2D skyrmion, toron, and uniform state. (a),(b),(c),(d) Structural stability diagrams of different material parameters dependent on the LC film thickness and surface anchoring, based on Frank-Oseen free energy minimization. Blue, red, and black squares indicate stable 2D skyrmion, toron, and uniform state. Filled (unfilled) squares indicate the uniform (TIC) state is the lower energy state, and a yellow (cyan) background indicates the solitons have lower (higher) energies than the uniform unwound state. (e),(f),(g) Mid-plane cross-sections of a unit cell of a stable toron lattice in a plane perpendicular (e) and horizontal (f,g) to the far field. The point defects are marked by red circles. (h) Preimages of points on $\mathbb{S}^2$ of an individual toron in the lattice indicated by cones in the top-right inset. (i),(j),(k) Mid-plane cross-sections of a unit cell of a stable skyrmion lattice in a plane perpendicular (i) and horizontal (j,k) to the far field. The point defects are marked by red circles. (l) Preimages of points on $\mathbb{S}^2$ of an individual skyrmion in the lattice indicated by cones in the top-right inset.

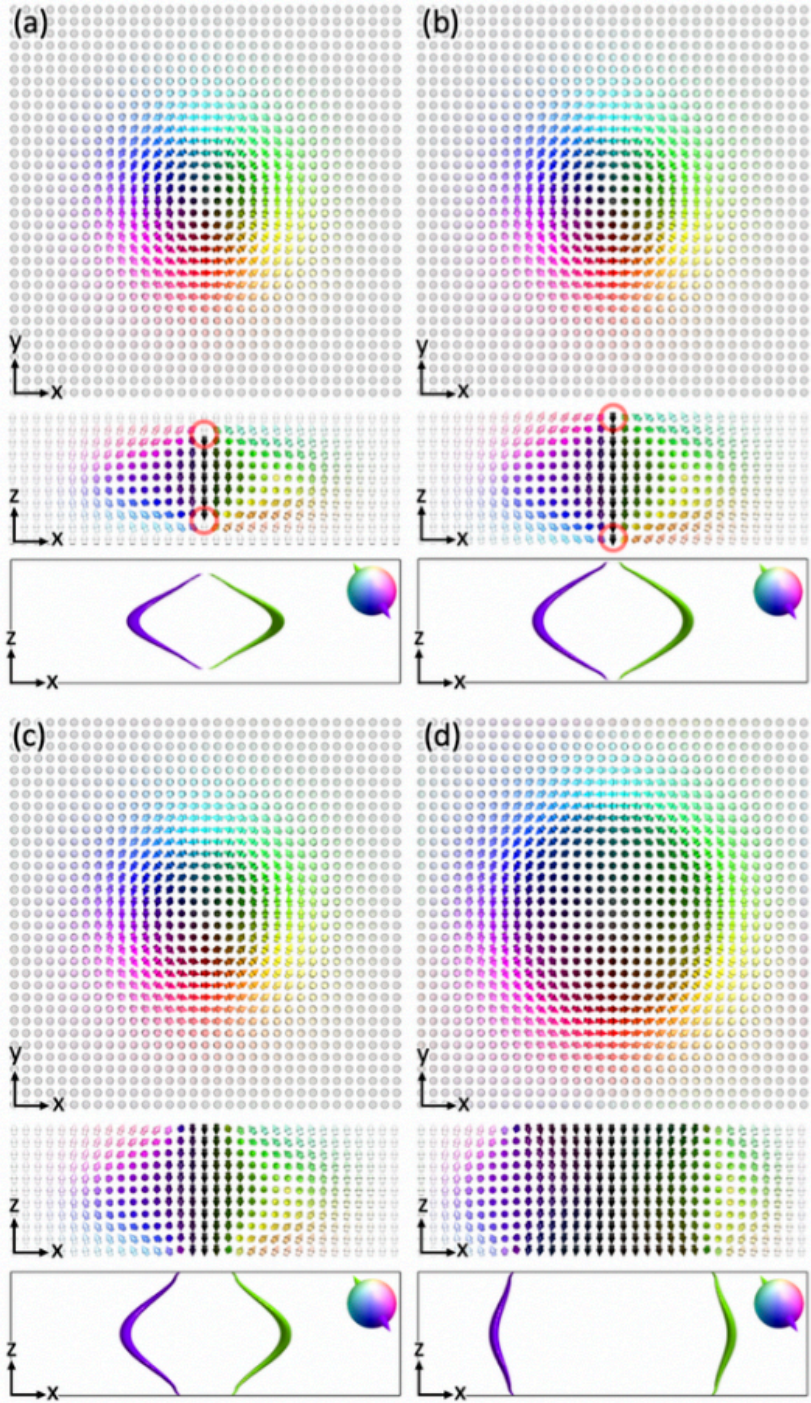

FIG. 4. Escape of point defects under weak anchoring conditions. (a) to (d) Snapshots of the point defects of a toron escaping beyond the surfaces and the toron transforming into a skyrmion when the anchoring condition is softened. Midplane cross-sections in a plane perpendicular (parallel) to the far field are shown in the top (middle) panels. The bottom panels show preimages of each structure of points on $\mathbb{S}^2$ indicated by cones in the top-right inset.

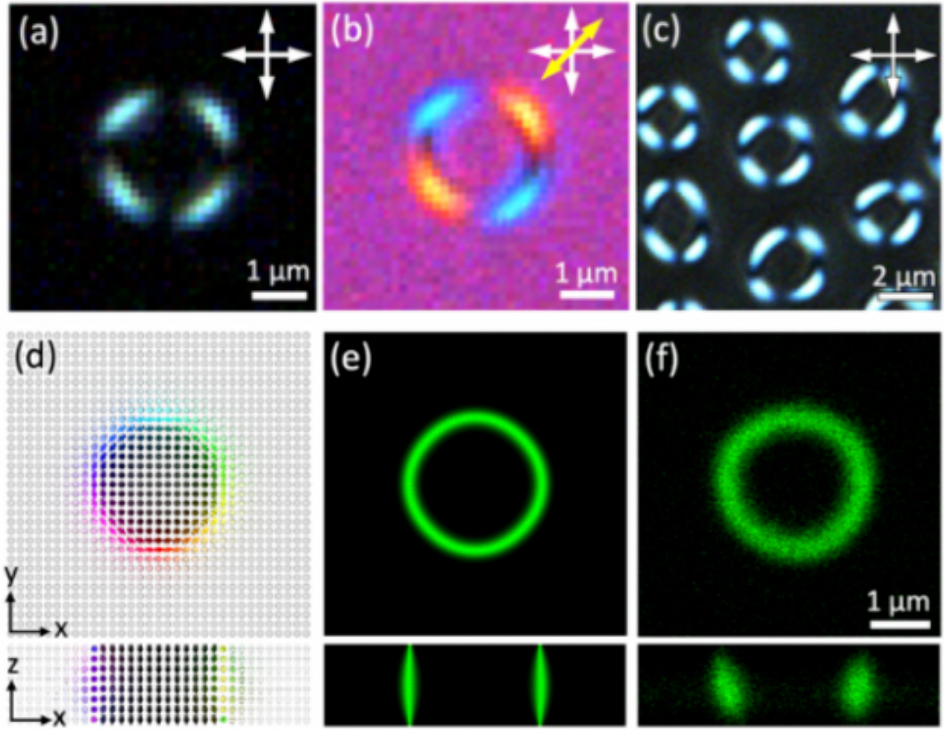

FIG. 5. 2D skyrmion. (a),(b) Experimental polarizing optical micrographs of a 2D skyrmion obtained without (a) and with (b) a 530-nm phase retardation plate with its slow axis labeled by the yellow double arrow. (c) Experimental polarizing optical micrograph of an assembly of 2D skyrmions. (d) Mid-plane cross-sections of a numerically simulated stable skyrmion in a plane perpendicular (top) and parallel (bottom) to the far field. (e) Numerically simulated 3PEF-PM images obtained with circular polarization of a skyrmion in (d) in the mid-planes perpendicular (top) and parallel (bottom) to the far field. (f) Experimental 3PEF-PM images obtained with circular polarization of a skyrmion in (a) and (b) in the mid-planes perpendicular (top) and parallel (bottom) to the far field. $d = 0.8$ μm and $p \sim 1$ μm.

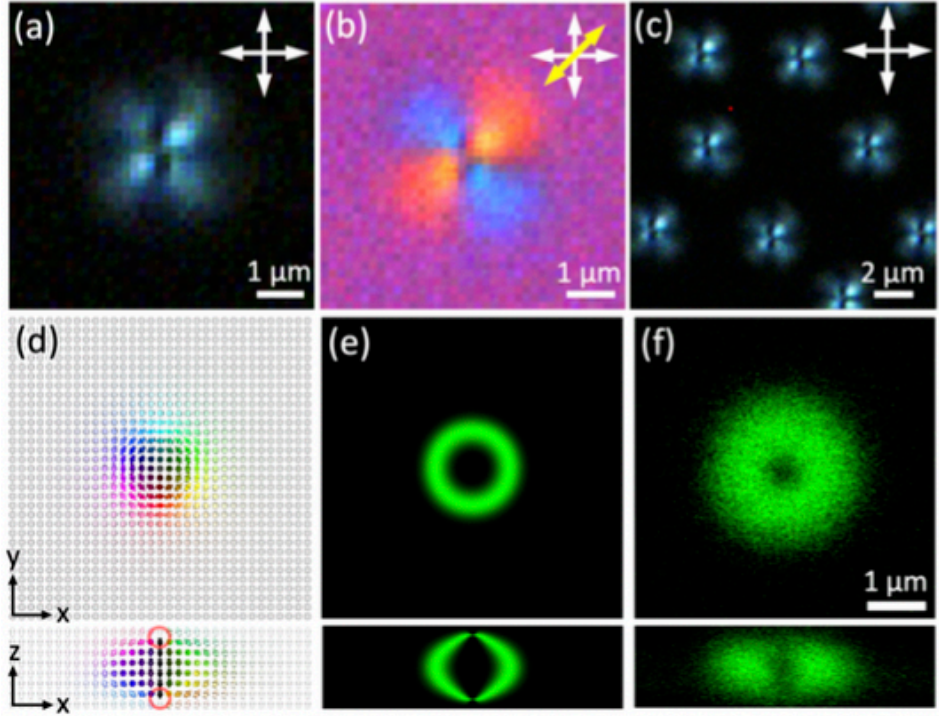

FIG. 6. Toron. (a),(b) Experimental polarizing optical micrographs of a toron obtained without (a) and with (b) a 530-nm phase retardation plate with its slow axis labeled by the yellow double arrow. (c) Experimental polarizing optical micrograph of an assembly of multiple torons. (d) Mid-plane cross-sections of a numerically simulated stable toron in a plane perpendicular (top) and parallel (bottom) to the far field. (e) Numerically simulated 3PEF-PM images obtained with circular polarization of a toron in (d) in the mid-planes perpendicular (top) and parallel (bottom) to the far field. (f) Experimental 3PEF-PM images obtained with circular polarization of a toron in (a) and (b) in the mid-planes perpendicular (top) and parallel (bottom) to the far field. $d = 1.05$ μm and $p = 1$ μm.

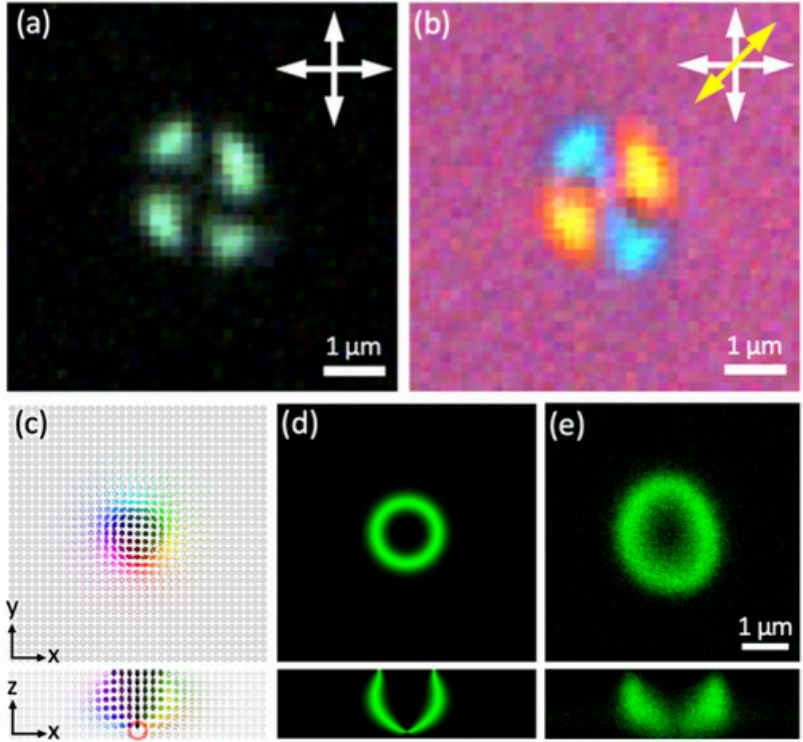

FIG. 7. Skyrmion/toron hybrid. (a),(b) Experimental polarizing optical micrographs of a skyrmion/toron hybrid structure obtained without (a) and with (b) a 530-nm phase retardation plate with its slow axis labeled by the yellow double arrow. (c) Mid-plane cross-sections of a numerically simulated stable skyrmion/toron hybrid in a plane perpendicular (top) and parallel (bottom) to the far field. (d) Numerically simulated 3PEF-PM images obtained with circular polarization of a skyrmion/toron hybrid in (c) in the mid-planes perpendicular (top) and parallel (bottom) to the far field. (e) Experimental 3PEF-PM images obtained with circular polarization of a skyrmion/toron hybrid in (a) and (b) in the mid-planes perpendicular (top) and parallel (bottom) to the far field. $d = 1$ µm and $p = 1$ µm.

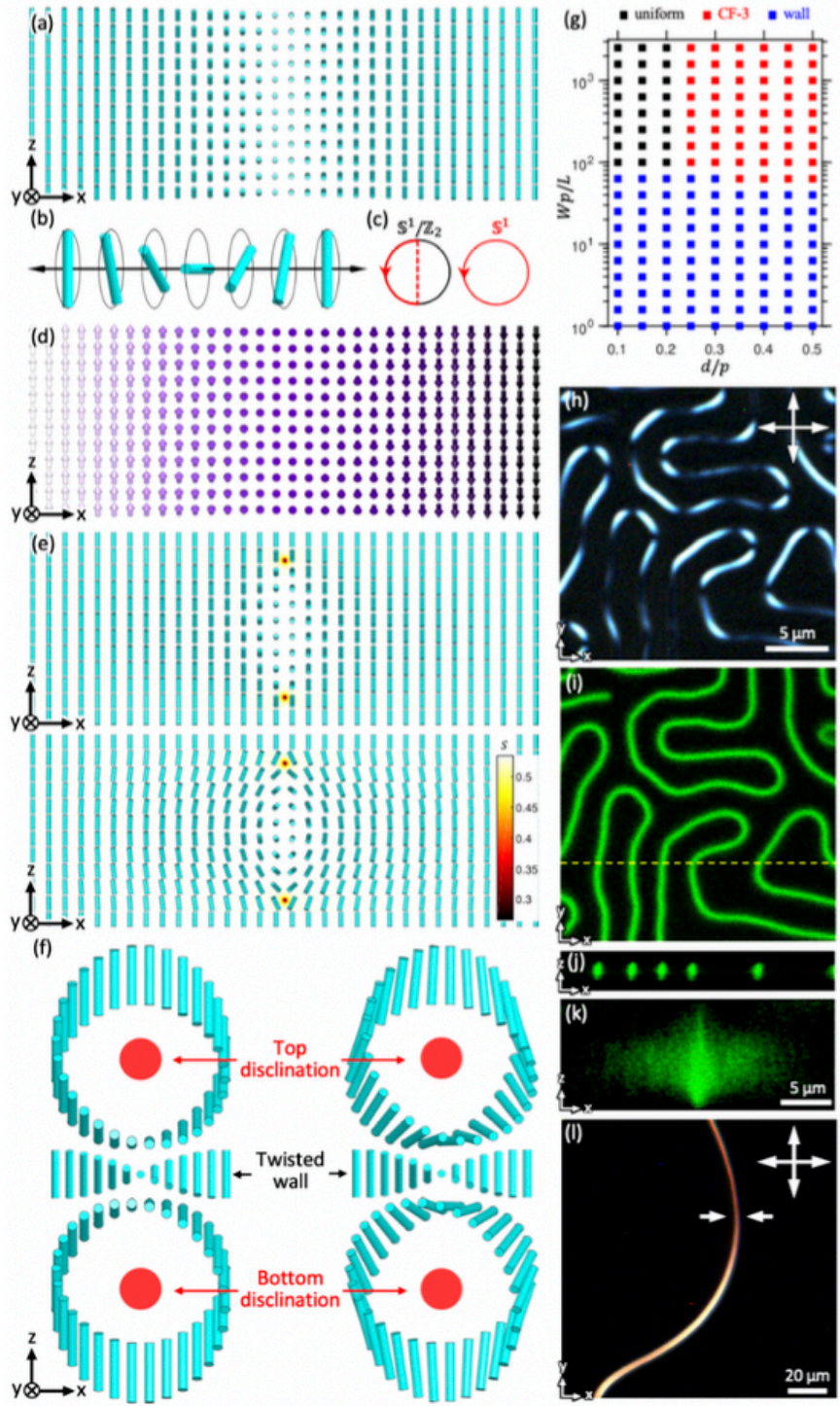

FIG. 8. Twisted wall and CF-3. (a) A LC twisted wall. (b) Schematic of a twisted wall of directors with head-tail symmetry going through a full $\pi$ rotation. (c) The director of the twisted wall in (b) winds around its order-parameter space $\mathbb{S}^1/\mathbb{Z}_2$ exactly once and $\mathbb{S}^1/\mathbb{Z}_2 \cong \mathbb{S}^1$. (d) The twisted wall in (a) with the field vectorized and visualized by arrows. (e) A translationally invariant CF-3 along the direction perpendicular to the cross-section with strong boundary conditions with (top) the 2D director orientations confined in the yz-plane and (bottom) 3D director orientations. The two half

integer defect lines near the substrates are visualized by reduction of $S$ at the defect cores. (f) Detailed field configurations visualized along circular loops of radius of 20 nm around the cores of the top and bottom singular disclinations with (left) 2D director orientations confined in the yz-plane and (right) unconfined 3D director orientations. (g) Structural stability diagram of twisted wall, CF-3, and uniform state dependent on the LC film thickness over pitch ratio and effective anchoring strength. $L$ is the elastic constant related to the average Frank elastic constant as $L = 2K/9S_{eq}^2$. (h) Experimental polarizing optical micrograph of twisted walls in a thin cell with $d = 0.8$ µm and $p = 1$ µm. (i),(j) Experimental 3PEF-PM images obtained with circular polarization of twisted walls in (h) in the mid-plane perpendicular to the far field (i) and in the plane parallel to the far field (j) through the yellow dashed line in (i). (k) Experimental 3PEF-PM image obtained with circular polarization of a CF-3 in the vertical plane between the arrows in (l). (l) Experimental polarizing optical micrograph of a CF-3. In (k) and (l), $d = 10$ µm and $p = 12.5$ µm.

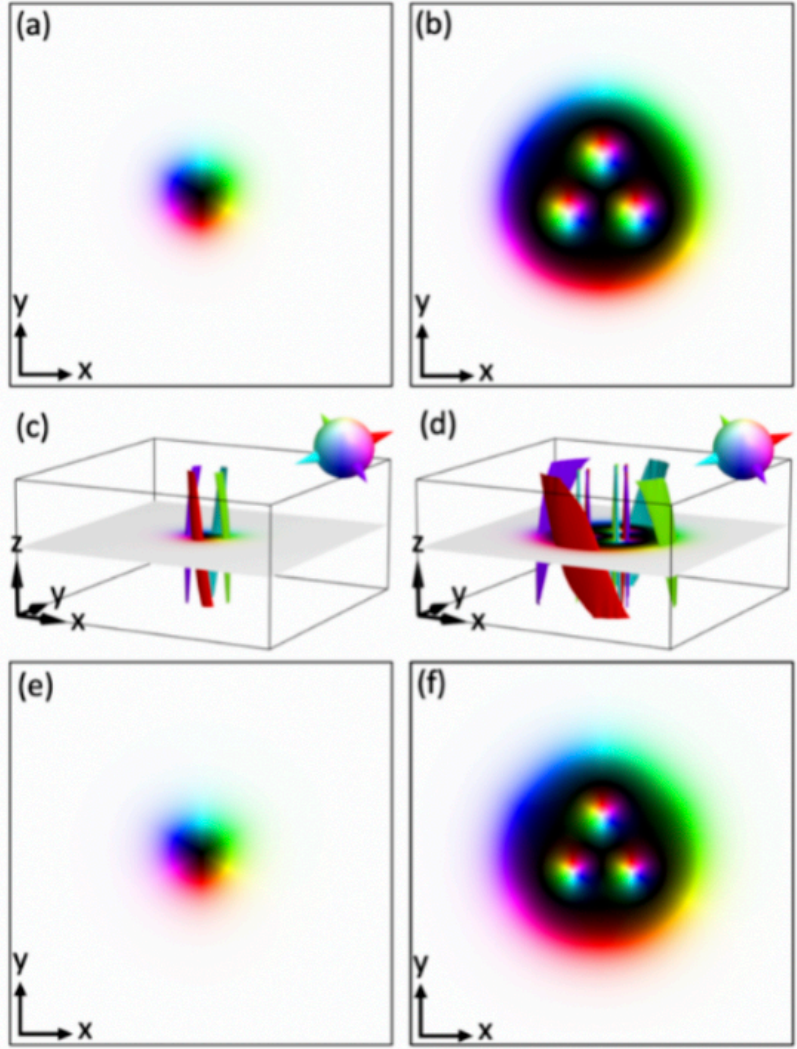

FIG. 9. 2D and 3D simulations of soliton structures. (a),(b) 2D simulations of an elementary skyrmion and a high charge skyrmion bag ($|N_{sk}| = 2$) visualized by colors based on the orientation of the vectorized field [Fig. 1(a), inset]. (c),(d) 3D simulations of an elementary skyrmion and a high charge skyrmion bag, respectively, shown by preimages of points on $\mathbb{S}^2$ indicated by cones in the top-right inset and the mid-plane cross-sections. (e),(f) The mid-plane cross-sections in (c) and (d) of an elementary skyrmion and a skyrmion bag, respectively.

# Tables:

**Table 1.** Material parameters of nematic LCs.

| Nematic host material | $K_{11}$ (pN) | $K_{22}$ (pN) | $K_{33}$ (pN) | $K$ (pN) | $\Delta n$ |
|---|---|---|---|---|---|
| 5CB | 6.4 | 3 | 10 | 6.47 | 0.18 |
| E7 | 10.8 | 6.8 | 17.5 | 11.7 | 0.23 |
| ZLI-3412 | 14.1 | 6.7 | 15.5 | 12.1 | 0.078 |
| ZLI-2806 | 14.9 | 7.9 | 15.4 | 12.4 | 0.04 |
| AMLC-0010 | 17.2 | 7.51 | 17.9 | 14.2 | 0.08 |